\documentclass[10pt,a4paper]{article}
\usepackage{authblk}
\usepackage[utf8]{inputenc}
\usepackage{amssymb}
\usepackage{amsmath}
\usepackage{empheq}
\usepackage{textcomp}
\usepackage{gensymb}
\usepackage{geometry}
\usepackage{dsfont}
\usepackage{mathrsfs}
\usepackage{arydshln}
\usepackage{relsize}
\usepackage{float}
\usepackage{graphicx}
\usepackage{csquotes}
\usepackage{mathdots}

\usepackage{tikz}
\usetikzlibrary{positioning,arrows}
\usetikzlibrary{decorations.pathmorphing}
\usetikzlibrary{decorations.markings}
\usetikzlibrary{shapes.geometric}
\usepackage{pgflibrarysnakes}
\tikzset{
    vector/.style={decorate, decoration={snake}, draw},
    provector/.style={decorate, decoration={snake,amplitude=2.5pt}, draw},
    antivector/.style={decorate, decoration={snake,amplitude=-2.5pt}, draw},
    fermion/.style={draw=black, postaction={decorate},decoration={markings,mark=at position .55 with {\arrow[draw=black]{>}}}},
    fermionbar/.style={draw=black, postaction={decorate},
                       decoration={markings,mark=at position .55 with {\arrow[draw=black]{<}}}},
    fermionnoarrow/.style={draw=black},
    graviton/.style={decorate, draw=black,decoration={coil,amplitude=4pt, segment length=5pt}},
    scalar/.style={dashed,draw=black, postaction={decorate},decoration={markings,mark=at position .55 with {\arrow[draw=black]{>}}}},
    scalarbar/.style={dashed,draw=black, postaction={decorate},decoration={markings,mark=at position .55 with {\arrow[draw=black]{<}}}},
    scalarnoarrow/.style={dashed,draw=black},
    electron/.style={draw=black, postaction={decorate},decoration={markings,mark=at position .55 with {\arrow[draw=black]{>}}}},
    bigvector/.style={decorate, decoration={snake,amplitude=4pt}, draw},
}

\usepackage{geometry}
\usepackage{hyperref}
\hypersetup{
  colorlinks   = true, 
  urlcolor     = blue, 
  linkcolor    = blue, 
  citecolor   = red 
}

\begin{document}

\begin{titlepage}
 
 \begin{center}
 \hfill
 \\	[22mm]
 
  {\Huge{Double Soft Theorem for Perturbative Gravity}}\\
  \vspace{16mm}
  
\large {Arnab Priya Saha}\\
\vspace{5mm}
{ \it Institute of Mathematical Sciences, \\ }
{ \it C.I.T Campus, Taramani, Chennai 600113, India\\}
and\\
{ \it Homi Bhabha National Institute, \\}
{ \it Training School Complex, Anushakti Nagar, Mumbai 400085, India.\\ }
\vspace{2mm}
email: \href{mailto: arnabps@imsc.res.in}{arnabps@imsc.res.in}

\end{center}

\medskip
\vspace{40mm}

\begin{abstract}

Following up on the recent work of Cachazo, He and Yuan \cite{DoubleSoftPRD}, we derive the double soft graviton theorem in perturbative gravity. We show that the double soft theorem derived using CHY formula precisely matches with the perturbative computation involving Feynman diagrams. In particular, we find how certain delicate limits of Feynman diagrams play an important role in obtaining this equivalence.

\end{abstract}

\end{titlepage}

\section{Introduction}
In a series of remarkable papers \cite{Cachazo:KLT, Cachazo:graviton, Cachazo:YM, PRL}, Cachazo et al have proposed a  formula for tree-level 
scattering amplitudes involving massless particles in any dimension. By means of a beautiful observation that the kinematic space of scattering 
data  involving $ n $ particles (namely Mandelstam type variables) can be mapped onto moduli space of $ n $-punctured Riemann sphere, they mapped 
the complicated problem of expressing tree-level scattering amplitude in terms of (an exponentially large number of) Feynman graphs into evaluating 
some relatively simple integrals over the moduli space. This formulation is likely to have serious ramifications for our understanding of quantum field 
theory and their dependence on space-time.

One of the elegant corollaries for understanding tree level amplitude in this light has been a new class of factorization theorems called 
the double soft theorems. In \cite{DoubleSoftPRD} the authors showed that, given a CHY formula for scattering amplitude one can look at limits in 
which two of the particles (gauge bosons or gravitons) become soft. Precisely as in the case of soft theorems like Weinberg's soft theorems where 
one of the massless gauge bosons becomes soft, CHY's analysis shows that the even in the double soft limit, the scattering amplitude factorizes. 
Cachazo et al derived such double soft theorems for a host of theories. However unlike in single soft case, double soft factors are more involved functions of momenta and
polarization. As such it is not immediately clear which of the Feynman diagrams contribute to the soft factor.

What is more significant is the fact that as the CHY formulation of gravitational scattering amplitude does not explicitly refer to any Lagrangian, 
it is a non-trivial check for the formula, if indeed the double soft theorem, which would probe three and four point vertices of Einstein
Hilbert action can be understood precisely via Feynman diagrammatic. In this paper, we apply the seminal ideas of \cite{DoubleSoftPRD} to CHY
formula for scattering amplitudes in perturbative gravity, thereby obtaining a double soft theorem for gravity scattering amplitude. The
formula looks rather formidable, but as we show, it precisely corresponds to the double soft limit of scattering amplitude obtained from EH action. 
The remarkable thing about the CHY formula is that, not only does it account for the two gravitons going soft at the same rate, it also has a contribution from
Feynman diagrams, where one graviton becomes soft at a faster rate than the second one. 

Recently double soft limits of scattering amplitudes have been explored for large variety of theories. In \cite{Volovich:string} double soft theorems for Yang Mills, supersymmetric gauge theories and open superstring theory have been studied. Single and consecutive double soft limits of gluon and graviton amplitudes have been analyzed in \cite{Klose:gluon}. Double soft theorems have also been studied in supergravity theories in \cite{Chen:sugra, He:twistorstring}. Studies of soft theorems are useful in uncovering the hidden symmetries of quantum field theories. In \cite{Strominger:Scat} connection between Weinberg's soft theorem and Ward identities of BMS symmetries at null infinity was shown. In \cite{Low:NLSM} it was shown that double soft theorems in nonlinear sigma model follow from a shift symmetry. It will be interesting to investigate the symmetries, if they exist, in relation to the double soft limit of gravity. This motivates us to look for a compact expression for gravity scattering amplitude in the double soft limit. As we will see double soft factor is not just product of two single soft factors but it also contains complicated factors at the sub-leading order.  

The paper is organized as follows. In sec(\ref{sec:CHY scatt}) we summarize the general philosophy of CHY formula of scattering amplitude and how to 
get single soft limit in it. In sec(\ref{sec:Double Soft}) we discuss the double soft formula in general. In sec(\ref{sec:EM DS}) we
simplify the double soft formula for Einstein Maxwell theory given in \cite{Cachazo:supp}. As a warm up exercise for calculating gravity amplitude 
from Feynman diagrams we do the same for Einstein-Maxwell which matches exactly with the CHY result. In sec(\ref{sec:EH}) we present the result for double soft graviton. 
In sec(\ref{sec:Feynman gravity}) we compute the Feynman diagrams for linearized gravity.The result matches with the previous one. All the details of the
calculations are presented in the appendix.

\section{A Brief Review of CHY Formula and Weinberg Soft Theorem}
\label{sec:CHY scatt}

In this section we review the CHY formula for tree level scattering amplitude of gravity. Details of the general formalism involving massless particles 
can be found in \cite{Cachazo:KLT, Cachazo:graviton, Cachazo:YM, PRL, Yuan}. The essential feature of the formalism is to map the singularities of the scattering
amplitude in the kinematic space of say, $ n $ massless particles to the singularity structure of an auxiliary space which is better understood. In this case Cachazo 
et al consider the moduli space of all $ n $-punctured Riemann sphere, $ \mathbb{CP}^{1} $. Let $ \{k_{1}^{\mu}, k_{2}^{\mu}, \ldots k_{n}^{\mu}\} $ are 
the momenta of $ n $ massless particles in $ D $ dimension and $ \{\sigma_{1}, \sigma_{2}, \ldots \sigma_{n} \}$ are holomorphic variables which parametrize the moduli space.
The holomorphic variables specify the locations of points on the Riemann sphere. The mapping of the singularities is given by \cite{Cachazo:3dscatt}

\begin{equation}
 k_{a}^{\mu} = \frac{1}{2\pi}\oint_{|z - \sigma_{a}| = \varepsilon} \mathrm{d}z \frac{f^{\mu}(z)}{\prod\limits_{b=1}^{n}(z-\sigma_{b})}, \qquad \forall a \in \{1,2, \dots n\}
\end{equation}
where $ f^{\mu}(z) $ is a $ D $ degree $ n-2 $ polynomials. 

Using momentum conservation, $ \sum\limits_{a=1}^{n}k_{a}^{\mu} = 0$ and the fact that $ k_{a}^{2} = 0 $ a set of $ n $ equations, called \textit{scattering equations}, can be 
derived 

\begin{equation}\label{scatt eqn}
 \sum\limits_{\substack{b=1 \\ b\ne a}}^{n} \frac{k_{a}.k_{b}}{\sigma_{a}-\sigma_{b}} = 0, \qquad \forall a \in \{1,2,\ldots n\}.
\end{equation}
However because of the invariance of the equations under the  $ \mathbb{SL}(2,\mathbb{C})$ transformation 

\begin{equation}
 \sigma \rightarrow \frac{\alpha \sigma + \beta}{\gamma\sigma + \delta}, \qquad \alpha, \: \beta, \: \gamma, \: \delta \in \mathbb{C}, 
	      \quad \alpha\delta - \beta\gamma = 1
\end{equation}
$ n-3 $ equations are independent and we can fix the values of $ \sigma_{1} \rightarrow \infty $, $ \sigma_{2} \rightarrow 0 $ and $ \sigma_{3} \rightarrow 1 $.

\subsection{Scattering Amplitude}
\label{sec:scatt amplitude}

CHY formula proposes an integral representation of the scattering amplitude of massless particles at tree level using the scattering equations on the complex Riemann sphere. 
For scattering of $ n $ particles it is given by

\begin{equation}\label{int rep}
 M_{n} = \int \frac{\mathrm{d}^{n}\sigma}{\mathrm{vol}\mathbb{SL}(2\mathbb{C})} \sideset{}{'}\prod\limits_{a} \delta \left( \sum\limits_{b \ne a}\frac{k_{a}.k_{b}}
		      {\sigma_{a}-\sigma_{b}}\right) I_{n}(\{k,\epsilon,\sigma\})
\end{equation}
where vol$\mathbb{SL}(2,\mathbb{C})$ is given by $\frac{\mathrm{d}\sigma_{a}\mathrm{d}\sigma_{b}\mathrm{d}\sigma_{c}}{(\sigma_{a}-\sigma_{b})
(\sigma_{b}-\sigma_{c})(\sigma_{c}-\sigma_{a})} $ for any $ a, b , c $. The primed product is defined as 

\begin{equation}
 \sideset{}{'}\prod\limits_{a} \delta \left( \sum\limits_{b \ne a}\frac{k_{a}.k_{b}}{\sigma_{a}-\sigma_{b}}\right) := 
			  (\sigma_{i}-\sigma_{j})(\sigma_{j}-\sigma_{k})(\sigma_{k}-\sigma_{i})\prod\limits_{a\ne i,j,k}
			  \delta \left( \sum\limits_{b \ne a}\frac{k_{a}.k_{b}}{\sigma_{a}-\sigma_{b}}\right)
\end{equation}
for any $ i,j,k $. Under the $\mathbb{SL}(2,\mathbb{C})$ transformation it can be checked that Eq.(\ref{int rep}) remains invariant.

In case of gravity the integrand is expressed as a function called the \textit{reduced Pfaffian} of an antisymmetric matrix which contains information about the momenta and polarization tensors of 
the particles. The antisymmetric matrix is defined as

\begin{equation}
 \Psi_{n} = 
 \left(
 \begin{array}{c:c}
  A & -C^{T} \\
  \hdashline
  C^{T} & B
 \end{array}
 \right)
\end{equation}
where each of $ A, B $ and $ C $ is $ 2n\times 2n $ matrix and the components are:

\begin{equation}
\begin{aligned}[c]
A_{ab} =
\begin{cases}
\frac{k_{a}.k_{b}}{\sigma_{a} - \sigma_{b}}, & a \ne b \\
0, & a=b
\end{cases}
\end{aligned}
\qquad
\begin{aligned}[c]
B_{ab} = 
\begin{cases}
 \frac{\epsilon_{a}.\epsilon_{b}}{\sigma_{a} - \sigma_{b}}, & a \ne b \\
 0, & a=b
\end{cases}
\end{aligned}
\qquad
\begin{aligned}[c]
C_{ab} =
\begin{cases}
 \frac{\epsilon_{a}.k_{b}}{\sigma_{a}-\sigma_{b}}, & a \ne b\\
 -\sum\limits_{c \ne a}\frac{\epsilon_{a}.k_{c}}{\sigma_{a}-\sigma_{c}}, & a=b.
\end{cases}
\end{aligned}
\end{equation}
The Pfaffian of $ \Psi_{n} $ vanishes because it has a nontrivial kernel of dimension two, spanned by the vectors:

\begin{equation}
 (1,1, \ldots, 1; 0,0, \ldots, 0)^{T} \quad \text{and} \quad (\sigma_{1}, \sigma_{2}, \ldots, \sigma_{n}; 0,0, \ldots, 0)^{T}.
\end{equation}
 Hence a new quantity, called the \textit{reduced Pfaffian}, is used:

\begin{equation}
 \mathrm{Pf'}\Psi_{n} = \frac{(-1)^{i+j}}{(\sigma_{i} - \sigma_{j})} \mathrm{Pf}(\Psi_{n})^{ij}_{ij}, \qquad \text{for any $i,j \in \{1,2, \ldots n\} $}.
\end{equation}
$(\Psi_{n})^{ij}_{ij} $ means the matrix obtained by deleting $i$ and $ j $ th rows and columns from $ \Psi_{n} $. This quantity is independent of the choice of $ i $ and $ j $.
In terms of the reduced Pfaffian, integrand for gravity scattering amplitude proposed by CHY is

\begin{equation}\label{gravity integrand}
 I_{n} = \left(\mathrm{Pf'}\Psi_{n}(\{k,\epsilon,\sigma \})\right)^{2}
\end{equation}
and using Eq.(\ref{int rep}) the tree level scattering amplitude for gravity in CHY formula becomes

\begin{equation}\label{scat amp grav}
 M_{n} = \int \frac{\mathrm{d}^{n}\sigma}{\mathrm{vol}\mathbb{SL}(2\mathbb{C})} \sideset{}{'}\prod\limits_{a} \delta \left( \sum\limits_{b \ne a}\frac{k_{a}.k_{b}}
		      {\sigma_{a}-\sigma_{b}}\right) (\mathrm{Pf'}\Psi_{N}(\{k,\epsilon,\sigma \}))^{2}.
\end{equation}
In the soft limit where energy of one of the scattered gravitons tends to zero $M_{n}$ can be factorized as a product of soft factor and scattering amplitude of 
remaining $n-1$ particles, precisely giving Weinberg's soft graviton theorem \cite{Weinberg}.

\subsection{Single Soft Limit}
\label{sec:single soft}

Here we explore what happens when energy of one of the scattered particles goes to zero. Details of the calculation can be found in \cite{PRL, Yuan, Cachazo:singsupp, Schwab, Afkhami}.
Let us assume momentum of the $ n $th particle scales as $ \tau p $ in the limit $ \tau \rightarrow 0 $. Then the scattering equation (\ref{scatt eqn}) for the 
$ n $th particle 

\begin{equation}
 f_{n} = \tau \sum\limits_{b=1}^{n-1}\frac{p.k_{b}}{\sigma_{n} - \sigma_{b}}
\end{equation}
trivially tends to zero and we are left with $ n-1 $ equations out of which $ n-4 $ are independent. Thus $ \sigma_{n} $ corresponding to the soft particle has 
no solution and the delta function supported at the $ n $th scattering equation is used to deform the $ \sigma_{n} $ integral to a contour integration where the 
contour wraps the solutions to the scattering equations

\begin{equation}\label{cont int}
 \int\mathrm{d}\sigma_{n}\delta\left(\tau \sum\limits_{b=1}^{n-1}\frac{p.k_{b}}{\sigma_{n} - \sigma_{b}} \right) \rightarrow 
		  \oint\frac{\mathrm{d}\sigma_{n}}{2\pi i}\frac{\tau^{-1}}{\sum\limits_{b=1}^{n-1}\frac{p.k_{b}}{\sigma_{n} - \sigma_{b}}}.
\end{equation}
Using a Taylor series expansion for other delta functions we get

\begin{equation}\label{delta}
 \sideset{}{'}\prod\limits_{a \ne n} \delta \left( \sum\limits_{b \ne a}\frac{k_{a}.k_{b}}{\sigma_{a}-\sigma_{b}}\right) = 
	      \sideset{}{'}\prod\limits_{a \ne n}\left[ \delta \left( \sum\limits_{b \ne a,n}\frac{k_{a}.k_{b}}{\sigma_{a}-\sigma_{b}}\right) + \tau\delta '\left(\frac{k_{a}.p}{\sigma_{a}-\sigma_{n}}\right) 
	      +\mathcal{O}(\tau^{2})\right].
\end{equation}
We use the Pfaffian expansion (\ref{Pf expansion}) to write 

\begin{equation}\label{sing Pfaff}
 \mathrm{Pf'}(\Psi_{n}) = \mathrm{Pf'}(\Psi_{n-1})\sum\limits_{b=1}^{n-1}\frac{\epsilon_{n}.k_{b}}{\sigma_{n}-\sigma_{b}} + \mathcal{O}(\tau). 
\end{equation}
Now substituting equations (\ref{cont int}), (\ref{delta}) and (\ref{sing Pfaff}) in Eq.(\ref{scat amp grav}) and performing the contour integration the 
scattering amplitude for gravity in the single soft limit can be expressed as

\begin{equation}\label{sing soft grav}
 M_{n} = \left(\frac{1}{\tau}\sum\limits_{b=1}^{n-1}\frac{(\epsilon_{n}.k_{b})^{2}}{p.k_{b}} \right) M_{n-1} + \mathcal{O}(1)
\end{equation}
which is precisely the expression given by Weinberg \cite{Weinberg}. Here we use the convention that all the momenta are outgoing. The Feynman diagram for linearized gravity corresponding to the amplitude 
(\ref{sing soft grav}) is given below:

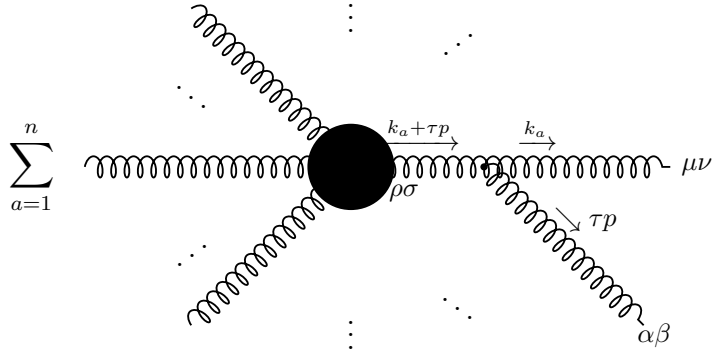
\begin{figure}[H]
 \begin{center}
  \begin{tikzpicture}[line width=0.7 pt, scale=0.7 pt]
  \node at (-6,0) {$\mathlarger{\mathlarger{\sum\limits_{a=1}^{n}}}$};
   \filldraw[] (0,0) circle(0.8);
   \draw[graviton] (-5,0) -- (0,0);
   \draw[graviton] (-3,3) -- (0,0);
   \draw[graviton] (-3,-3) -- (0,0);
   \draw[graviton] (0,0) -- (1,0)node[below=0.1]{$\rho\sigma$} -- (1.3,0)node[above=0.1]{$\xrightarrow{k_{a}+\tau p} $} -- (3.5,0)node[above=0.1]
			{$\xrightarrow{k_{a}} $} -- (6,0);
    \draw[graviton] (2.5,0) -- (3.5,-1)node[right=0.1]{$\searrow \tau p$} -- (5.5,-3);
    \filldraw[] (2.5,0) circle(0.05);
    \node at (6.5,0) {$\mu\nu$};
    \node at (5.7,-3.2) {$\alpha\beta$};
    \node at (-3,-1.5) {$\mathlarger{\mathlarger{\iddots}}$};
   \node at (-3,1.5) {$\mathlarger{\mathlarger{\ddots}}$};
   \node at (0,3) {$\mathlarger{\mathlarger{\vdots}}$};
   \node at (0,-3) {$\mathlarger{\mathlarger{\vdots}}$};
   \node at (2,-2.5) {$\mathlarger{\mathlarger{\ddots}}$};
   \node at (2,2.5) {$\mathlarger{\mathlarger{\iddots}}$};
  \end{tikzpicture}
 \end{center}
\end{figure}


\section{Double Soft Limit of CHY Formula}
\label{sec:Double Soft}

In this section we briefly summarize the analysis for the double soft limit of the CHY scattering amplitude as formulated by 
Cachazo et al \cite{DoubleSoftPRD, Cachazo:supp}. Let us consider a scattering process of massless particles with $ N=n+2 $ external legs, out of 
which momenta of two of the particles labeled by $ n+1 $ and $ n+2 $ are taken to be soft. The soft momenta are denoted by 

\begin{equation}
 k_{n+1} = \tau p, \qquad k_{n+2} = \tau q
\end{equation}
with the limit $ \tau \rightarrow 0 $. The variables $ \sigma_{n+1} $ and $ \sigma_{n+2} $ corresponding to the soft particles parametrized by new variables as 

\begin{equation}\label{sigma variables}
 \sigma_{n+1} = \rho - \frac{\xi}{2}, \qquad \sigma_{n+2} = \rho + \frac{\xi}{2}.
\end{equation}
In terms of the new variables the scattering equations (\ref{scatt eqn}) now become

\begin{equation}\label{scattering eqn}
 f_{a} = \begin{cases}
          \sum\limits_{\substack{b=1{}\\b\ne a}}^{n} \left(\frac{k_{a}.k_{b}}{\sigma_{a}-\sigma_{b}} + \frac{\tau k_{a}.p}{\sigma_{a}-\rho+\frac{\xi}{2}} 
          + \frac{\tau k_{a}.q}{\sigma_{a}-\rho-\frac{\xi}{2}}\right), & a \ne n+1, n+2\\
          \sum\limits_{b=1}^{n} \left(\frac{\tau k_{b}.p}{\rho-\frac{\xi}{2}-\sigma_{b}} - \frac{\tau^{2}p.q}{\xi}\right), & a = n+1\\
          \sum\limits_{b=1}^{n} \left(\frac{\tau k_{b}.p}{\rho+\frac{\xi}{2}-\sigma_{b}} + \frac{\tau^{2}p.q}{\xi}\right), & a = n+2.
         \end{cases}
\end{equation}
Expanding $ \xi $ perturbatively in $ \tau $ as\footnote{ we do not consider the non-degenerate solutions $(\xi \sim \tau^{0})$ because they contribute at
sub leading order of $ \tau $ compared to the degenerate solutions $(\xi \sim \tau)$ for most of the theories of interest \cite{DoubleSoftPRD}. For gravity we
show this argument explicitly in sec.(\ref{sec:EH}).}

\begin{equation}
 \xi = \tau \xi_{1} + \tau^{2}\xi_{2} + \mathcal{O}(\tau^{3})
\end{equation}
and using the last two scattering equations we get 

\begin{equation}\label{xi sol}
 \frac{1}{\xi_{1}} = \frac{1}{p.q}\sum\limits_{b=1}^{n}\frac{k_{b}.p}{\rho-\sigma_{b}} = - \frac{1}{p.q}\sum\limits_{b=1}^{n}\frac{k_{b}.q}{\rho-\sigma_{b}}.
\end{equation}
With the change of variables in (\ref{sigma variables}) the $ \sigma_{n+1} $ and $ \sigma_{n+2} $ integrals can be transformed as follows:

\begin{eqnarray}\label{DS delta}
 && \int\mathrm{d}\sigma_{n+1}\mathrm{d}\sigma_{n+2}\delta(f_{n+1})\delta(f_{n+2}) \nonumber \\
 & \rightarrow & -2\int\mathrm{d}\rho\mathrm{d}\xi\delta(f_{n+1}+f_{n+2})\delta(f_{n+1}-f_{n+2}) \nonumber\\
 & \rightarrow & -2\oint\frac{\mathrm{d}\rho}{2\pi i}\sum\limits_{\xi\: \text{solutions}}\int\mathrm{d}\xi \frac{1}{(f_{n+1} + f_{n+2})}
		  \frac{1}{\frac{\partial}{\partial \xi}(f_{n+1} - f_{n+2})} \nonumber \\
 & \rightarrow & -2\oint\frac{\mathrm{d}\rho}{2\pi i}\sum\limits_{\xi\: \text{solutions}}\int\mathrm{d}\xi \frac{1}{\sum\limits_{a=1}^{n}
		\tau\left(\frac{k_{a}.p}{\rho-\frac{\xi}{2}-\sigma_{a}}+ \frac{k_{a}.q}{\rho+\frac{\xi}{2}-\sigma_{a}} \right)}
		\frac{2}{\sum\limits_{b=1}^{n}\tau\left(\frac{k_{b}.p}{(\rho-\frac{\xi}{2}-\sigma_{b})^{2}}+ \frac{k_{b}.q}{(\rho+\frac{\xi}{2}-\sigma_{b})^{2}}\right)
		+ \frac{4 \:\tau^{2}\: p.q}{\xi^{2}}}
 \end{eqnarray}
 where the first delta constraint is expressed as a contour integral for $ \rho $ wrapping around the solutions to the scattering equations and the second delta 
 constraint localizes the $ \xi $ variable. It is evident that the scaling of the expression (\ref{DS delta}) goes as $ \frac{1}{\tau}$ if $ \xi \sim \tau $ and 
 it is $ \frac{1}{\tau^{2}} $ if $ \xi \sim \tau^{0} $.
 
For finite $ \rho $ contour the CHY expression for the scattering amplitude at tree level in the double soft limit as an expansion in the order of $ \tau $ is given by 
\cite{Cachazo:supp}

\begin{equation}
 M_{N} = -\frac{1}{\tau}\oint \frac{\mathrm{d}\rho}{2\pi i}\int\mathrm{d}\mu_{n}\frac{\xi_{1}^{2}}{p.q \sum\limits_{b=1}^{n}\frac{k_{b}.(p+q)}{\rho-\sigma_{b}}}
	  \left(1 - \frac{\tau\xi_{1}}{2} \frac{\sum\limits_{b=1}^{n}\frac{k_{b}.(p+q)}{(\rho-\sigma_{b})^{2}}}{\sum\limits_{b=1}^{n}
	  \frac{k_{b}.(p+q)}{\rho-\sigma_{b}}} + 3\tau\frac{\xi_{2}}{\xi_{1}} + \mathcal{O}(\tau^{2}) \right)I_{N}.
\end{equation}
Here we use the notation
$\mathrm{d}\mu_{n} \equiv \frac{\mathrm{d}^{n}\sigma}{\mathrm{vol}\mathbb{SL}(2\mathbb{C})}\sideset{}{'}\prod\limits_{a} \delta \left( \sum\limits_{b \ne a}\frac{k_{a}.k_{b}}{\sigma_{a}-\sigma_{b}}\right) $. If the integrand can be written as a product like

\begin{equation}
 I_{N}(k,\sigma,\rho,\xi) = F(k,\sigma,\rho,\xi)I_{n}(k,\sigma) + \text{(sub-leading order)}
\end{equation}
then the previous expression at  leading order simplifies to 

\begin{equation}\label{double soft factorization}
 M_{N} = \left[-\frac{1}{\tau}\oint \frac{\mathrm{d}\rho}{2\pi i}\frac{\xi_{1}^{2}}{p.q \sum\limits_{b=1}^{n}\frac{k_{b}.(p+q)}{\rho-\sigma_{b}}}F(k,\sigma,\rho,\tau\xi_{1})\right] M_{n}.
\end{equation}
The term in the square bracket gives the leading order double soft factor $ S^{*}(0) $.

There is an additional contribution to $ M_{N} $ coming from the pole at $ \rho = \infty $. Deforming the contour around the pole at infinity, the 
leading order expression can be derived to be

\begin{equation}\label{Pole inf}
 (M_{N})_{\infty} = \oint \frac{\mathrm{d}\rho}{2\pi i} \int\mathrm{d}\mu_{n} \frac{-2\rho^{-3}}{3\tau^{4}(p.q)^{2}}\left(I_{N}|_{\xi = 2 i \rho} 
		    + I_{N}|_{\xi = -2 i \rho}\right).
\end{equation}


\section{Double Soft Limit for Einstein Maxwell Theory}
\label{sec:EM DS}

Now we will like show how the double soft theorem follows from Feynman diagrams. As an example we consider Einstein Maxwell theory. In \cite{Cachazo:supp} the authors have investigated scattering amplitudes in Born Infeld and Einstein Maxwell theories in the double soft limit 
with two soft photons. The integrand for this class of theories is given by

\begin{equation}\label{EM CHY}
 I_{N} = (\mathrm{Pf}X_{N})^{-m} (\mathrm{Pf'}A_{N})^{2+m} \mathrm{Pf'}\Psi_{N}
\end{equation}
where $ m = 0, -1 $ denote BI and EM respectively. The result for EM theory with two soft photon emission is 

\begin{equation}
 S^{*(0)} = \frac{1}{\tau} \sum\limits_{b=1}^{n} \frac{1}{k_{b}.(p+q)} \left[ \frac{p.q \: \epsilon_{n+1}.\epsilon_{n+2} - \epsilon_{n+2}.p \: \epsilon_{n+1}.q}{4 (p.q)^{2}}
	    \bigl\{k_{b}.(p-q)\bigr\}^{2} - \epsilon_{n+1}.p_{b}^{\bot} \: \epsilon_{n+2}.q_{b}^{\bot} \right]
\end{equation}
which can be further simplified to\footnote{Here we have used $\sum\limits_{b=1}^{n}\frac{\{k_{b}.(p-q)\}^{2}}{k_{b}.(p+q)} = -4\sum\limits_{b=1}^{n}
\frac{k_{b}.p\: k_{b}.q}{k_{b}.(p+q)} + \mathcal{O}(\tau)$.} 

\begin{equation}\label{EM DS}
 S^{*(0)} = \frac{1}{\tau} \sum\limits_{b=1}^{n} \frac{1}{k_{b}.(p+q)} \left[ \frac{\epsilon_{n+1}.q \: \epsilon_{n+2}.k_{b} \: p.k_{b} 
	     + \epsilon_{n+1}.k_{b} \: \epsilon_{n+2}.p \: q.k_{b} - \epsilon_{n+1}.\epsilon_{n+2} \: p.k_{b} \: q.k_{b}}{p.q} - \epsilon_{n+1}.k_{b} \: \epsilon_{n+2}.k_{b} \right].
\end{equation}

In the following subsection we show that from Feynman diagrams we can reproduce the above expression modulo an overall constant factor.

\subsection{Feynman Diagrams }
\label{sec:FD for EM}

The Einstein Maxwell action in four dimension is given by

\begin{equation}\label{EM action}
 S_{EM} = \int\mathrm{d}^{4}x \left(-\frac{1}{4}\sqrt{-g}g^{\mu\rho}g^{\nu\sigma}F_{\mu\nu}F_{\rho\sigma} + \frac{2}{\kappa^{2}}\sqrt{-g}R\right)
\end{equation}
where $F_{\mu\nu} = \partial_{\mu}A_{\nu} - \partial_{\nu}A_{\mu} $ and $ R $ is the Ricci scalar given by $ R = g^{\mu\nu} \left(\Gamma^{\lambda}_{\mu\lambda\:,\nu}
-\Gamma^{\lambda}_{\mu\nu\:,\lambda} + \Gamma^{\sigma}_{\mu\lambda}\Gamma^{\lambda}_{\sigma\nu} - \Gamma^{\sigma}_{\mu\nu}\Gamma^{\lambda}_{\sigma\lambda}\right) $.

In the linearized perturbative theory of gravity a small deviation $ h_{\mu\nu} $ around flat Minkowski spacetime is considered such as

\begin{equation}\label{g pert}
 g_{\mu\nu} = \eta_{\mu\nu} + \kappa h_{\mu\nu}.
\end{equation}

The Feynman rules for EM are given in sec(\ref{sec:EM Feynman}). At leading order two soft photons can be emitted from either an external graviton 
leg through an internal graviton propagator or from an external photon leg through a graviton propagator. Both the processes involve two three-point $AAg$ vertices. 
These give terms of $\mathcal{O}\left(\frac{\kappa^{2}}{\tau}\right)$. There also exists a four-point $AAgg$ vertex through which two soft photons can come from an external graviton leg. This vertex comes from a 
term in Lagrangian of the form $\sim hh\partial A \partial A$ and thus lead to the order of $\mathcal{O}(\tau)$ in the scattering amplitude. 
Therefore for our purpose of interest it suffices to compute the following Feynman diagrams:

\begin{itemize}
 \item \textbf{photons emitted from an external graviton}
 
 
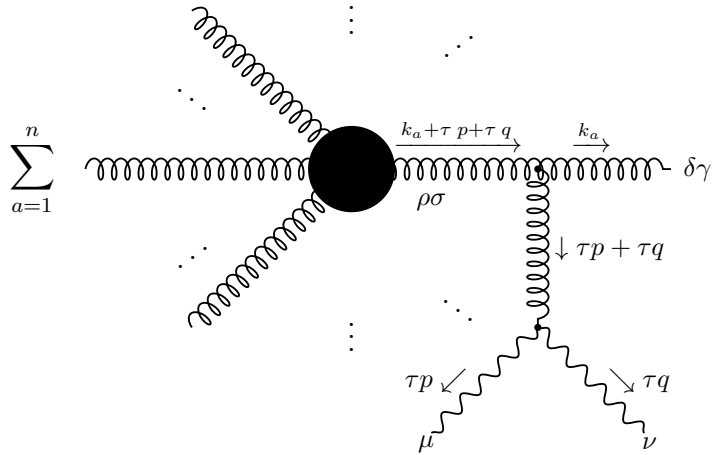
\begin{figure}[H]
\begin{center}
\begin{tikzpicture}[line width=0.7 pt, scale=0.7]
\node at (-6,0) {$\mathlarger{\mathlarger{\sum\limits_{a=1}^{n}}}$};
\filldraw[] (0,0) circle(0.8);
\draw[graviton] (-5,0) -- (0,0);
\draw[graviton] (-3,-3) -- (0,0);
\draw[graviton] (-3,3) -- (0,0);
\draw[graviton] (0,0) -- (1.5,0)node[below=0.2]{$\rho\sigma$} -- (2,0)node[above=.1]{$\xrightarrow{k_{a}+ \tau\:p+\tau\:q}$}  
		-- (4.5,0)node[above=0.1]{$ \xrightarrow{k_{a}}$} -- (6,0);
\node at (6.5,0) {$\delta\gamma$};
\draw[graviton] (3.5,0) --(3.5,-1.5)node[right=0.1]{$\downarrow \tau p+\tau q$}-- (3.5,-3);
\draw[vector] (3.5,-3) --(2.5,-4)node[left=0.1]{$ \tau p\swarrow$}-- (1.5,-5);
\draw[vector] (3.5,-3) -- (4.5,-4)node[right=0.1]{$ \searrow\tau q$} --(5.5,-5);
\filldraw[] (3.5,0) circle(0.05);
\filldraw[] (3.5,-3) circle(0.05);
\node at (1.4,-5.2) {$\mu$};
\node at (5.6,-5.2) {$\nu$};
\node at (-3,-1.5) {$\mathlarger{\mathlarger{\iddots}}$};
   \node at (-3,1.5) {$\mathlarger{\mathlarger{\ddots}}$};
   \node at (0,3) {$\mathlarger{\mathlarger{\vdots}}$};
   \node at (0,-3) {$\mathlarger{\mathlarger{\vdots}}$};
   \node at (2,-2.5) {$\mathlarger{\mathlarger{\ddots}}$};
   \node at (2,2.5) {$\mathlarger{\mathlarger{\iddots}}$};
\end{tikzpicture}
\end{center}
\end{figure}

\item \textbf{photons emitted from an external photon}

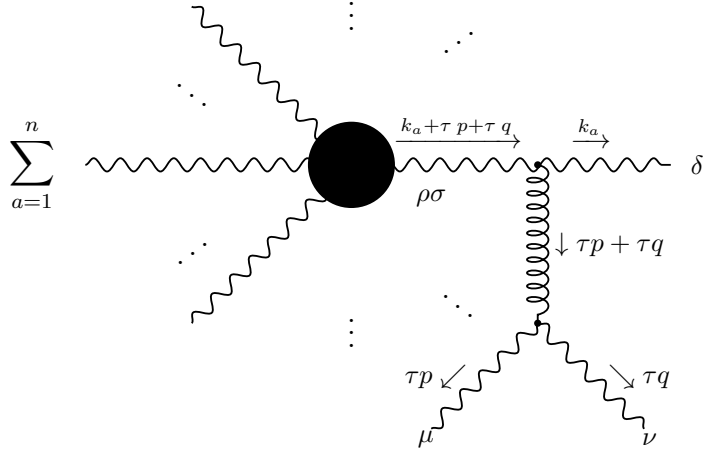
\begin{figure}[H]
\begin{center}
\begin{tikzpicture}[line width=0.7 pt, scale=0.7]
\node at (-6,0) {$\mathlarger{\mathlarger{\sum\limits_{a=1}^{n}}}$};
\filldraw[] (0,0) circle(0.8);
\draw[vector] (-5,0) -- (0,0);
\draw[vector] (-3,-3) -- (0,0);
\draw[vector] (-3,3) -- (0,0);
\draw[vector] (0,0) -- (1.5,0)node[below=0.2]{$\rho\sigma$} -- (2,0)node[above=.1]{$\xrightarrow{k_{a}+ \tau\:p+\tau\:q}$}  
		-- (4.5,0)node[above=0.1]{$ \xrightarrow{k_{a}}$} -- (6,0);
\node at (6.5,0) {$\delta$};
\draw[graviton] (3.5,0) --(3.5,-1.5)node[right=0.1]{$\downarrow \tau p+\tau q$}-- (3.5,-3);
\draw[vector] (3.5,-3) --(2.5,-4)node[left=0.1]{$ \tau p\swarrow$}-- (1.5,-5);
\draw[vector] (3.5,-3) -- (4.5,-4)node[right=0.1]{$ \searrow\tau q$} --(5.5,-5);
\filldraw[] (3.5,0) circle(0.05);
\filldraw[] (3.5,-3) circle(0.05);
\node at (1.4,-5.2) {$\mu$};
\node at (5.6,-5.2) {$\nu$};
\node at (-3,-1.5) {$\mathlarger{\mathlarger{\iddots}}$};
   \node at (-3,1.5) {$\mathlarger{\mathlarger{\ddots}}$};
   \node at (0,3) {$\mathlarger{\mathlarger{\vdots}}$};
   \node at (0,-3) {$\mathlarger{\mathlarger{\vdots}}$};
   \node at (2,-2.5) {$\mathlarger{\mathlarger{\ddots}}$};
   \node at (2,2.5) {$\mathlarger{\mathlarger{\iddots}}$};
\end{tikzpicture}
\end{center}
\end{figure}

\end{itemize}

Both of these diagrams give exactly the same soft factor as in Eq.(\ref{EM DS}). The above analysis demonstrates the correspondence between the CHY integrand (\ref{EM CHY}) with $m=-1$ and the Lagrangian description of EM theory (\ref{EM action}) in the 
double soft limit. This motivates us to look for similar kind of relation between CHY gravity proposal and Einstein gravity, which we explore in the next section.

\section{Double Soft Limit in Gravity Amplitude}
\label{sec:EH}
In this section we derive the double soft factor for gravity from CHY method. Details of the calculations are presented in sec(\ref{sec:appendix gravity integrand}, \ref{sec: app DS factor}, \ref{sec:app pole inf}). 
For pure gravity amplitude integrand is given by 

\begin{equation}
 I_{n+2} = (\mathrm{Pf'}\Psi_{n+2})^{2}.
\end{equation}
In the double soft limit for the degenerate solution $ (\xi \sim \tau) $, the measure goes as $ \mathrm{d}\mu_{n+2} \sim \tau^{-1} $ and the integrand 
goes as $ I_{n+2} \sim \tau^{0} $. So the overall scaling of the soft factor is $ \frac{1}{\tau} $. For the non-degenerate solution $ (\xi \sim \tau^{0}) $
the integrand will be $ I_{n+2} \sim \tau^{4} $ whereas, as argued following Eq.(\ref{DS delta}), the measure will scale as $\mathrm{d}\mu_{n+2} \sim \tau^{-2} $.
So the leading order of the double soft factor in case of non-degenerate solution of $\xi$ will be of $\mathcal{O}(\tau^{2})$. Hence like other theories \cite{DoubleSoftPRD} (sGal, DBI, 
EMS, NLSM, YMS) for EH gravity too the degenerate solution contributes at the leading order in the double soft factor. 

Through out the rest of the discussion we will consider the degenerate solution of $\xi$ and restrict to only the leading order term. In this case the integrand takes the form

\begin{equation}\label{grav ds Pf expansion}
 I_{n+2} = \left[\frac{\epsilon_{n+1}.q\:\epsilon_{n+2}.p - \epsilon_{n+1}.\epsilon_{n+2}\:p.q}{\xi_{1}^{2}} +\sum\limits_{i,j=1}^{n}
	    \frac{\epsilon_{n+1}.p_{i}^{\bot}\:\epsilon_{n+2}.q_{j}^{\bot}}{(\rho-\sigma_{i})(\rho-\sigma_{j})}\right]^{2}(\mathrm{Pf'}\Psi_{n})^{2} + \mathcal{O}(\tau).
\end{equation}
At leading order the double soft factor for gravity follows\footnote{There is another piece to this amplitude coming from Eq.(\ref{Pole inf}). We
show in Sec.(\ref{sec:app pole inf}) that contribution from pole at infinity is subleading to that of finite contour integral. Hence Eq.(\ref{double soft factorization})
is the leading order term.} from Eq.(\ref{double soft factorization}) and is given by

\begin{eqnarray}\label{gravity DS factor}
 S^{*(0)} &=& -\frac{1}{\tau}\sum\limits_{a=1}^{n}\left[ \frac{1}{k_{a}.(p+q) \: p.q}\biggl\{ -(\epsilon_{n+1}.\epsilon_{n+2})^{2}\: k_{a}.p\: k_{a}.q 
	      +2 \: \epsilon_{n+1}.\epsilon_{n+2} \left(\epsilon_{n+1}.q \:\epsilon_{n+2}.k_{a} \: k_{a}.p + \epsilon_{n+1}.k_{a} \: \epsilon_{n+2}.p \: k_{a}.q \right) \right. \nonumber \\
	      && \left. \phantom{-\frac{1}{\tau}\sum\limits_{a=1}^{n}\left[\frac{1}{k_{a}.(p+q) \: p.q}\biggl\{\right.} -2 \: \epsilon_{n+1}.q \: \epsilon_{n+2}.p \: \epsilon_{n+1}.k_{a} \: \epsilon_{n+2}.k_{a} 
	      + (\epsilon_{n+1}.q)^{2} \: (\epsilon_{n+2}.k_{a})^{2} + (\epsilon_{n+1}.k_{a})^{2}\:(\epsilon_{n+2}.p)^{2} \biggr\} \right. \nonumber \\
	      && \left. \phantom{-\frac{1}{\tau}\sum\limits_{a=1}^{n}\left[ \right.} - \frac{1}{p.q} \biggl\{ \frac{(\epsilon_{n+1}.q)^{2}\:(\epsilon_{n+2}.p)^{2}}{k_{a}.q} 
	      + \frac{(\epsilon_{n+1}.k_{a})^{2}\: (\epsilon_{n+2}.p)^{2}}{k_{a}.p} \biggr\} \right. \nonumber\\
	      && \left. \phantom{-\frac{1}{\tau}\sum\limits_{a=1}^{n}\left[ \right.} + \frac{1}{k_{a}.(p+q)} \biggl\{ -2 \: \epsilon_{n+1}.\epsilon_{n+2} \: \epsilon_{n+1}.k_{a} \:\epsilon_{n+2}.k_{a}
	      + 2 \: \epsilon_{n+1}.k_{a}\: \epsilon_{n+2}.k_{a} \left( \frac{\epsilon_{n+1}.q \: \epsilon_{n+2}.k_{a}}{k_{a}.q} +  \frac{\epsilon_{n+1}.k_{a} \: \epsilon_{n+2}.p}{k_{a}.p}\right) \right. \nonumber\\
	      && \left. \phantom{-\frac{1}{\tau}\sum\limits_{a=1}^{n}\left[ \frac{1}{k_{a}.(p+q)} \biggl\{ \right.} - \frac{(\epsilon_{n+1}.k_{a})^{2} \: (\epsilon_{n+2}.k_{a})^{2} \: p.q}{k_{a}.p \: k_{a}.q} \biggr\} \right]
\end{eqnarray}

\paragraph{Remarks on the double soft factor:}

We notice the following significant features about this soft factor:

\begin{itemize}
 \item It is interesting to see that like single soft factor (\ref{sing soft grav}), the double soft factor (\ref{gravity DS factor}) too appears at $\mathcal{O}\left( \frac{1}{\tau} \right)$ 
and not at $\mathcal{O}\left( \frac{1}{\tau^{2}} \right)$ as one might have expected. Although we will see later that Feynman diagrammatic give terms of $\mathcal{O}\left( \frac{1}{\tau^{2}} \right)$ which are not 
contained in the CHY formula. 

  \item It contains only summation over single variables. This is precisely
due to the fact that when we do contour integration only simple poles at each of the scattering solutions of $\sigma_{a}$, for $a \in \{1,2,3\ldots n\}$ contribute. From the perspective of 
Feynman diagrams this implies that CHY formula is capturing only those \textit{local} processes where two soft gravitons are emitted from 
the same external leg. In general soft gravitons can be scattered from different external legs, in that case summation over all such external legs 
have to be carried on. Absence of these terms implies that CHY formula in the double soft limit describes scattering processes which are \textit{local}.

  \item The gauge invariance of the expression (\ref{gravity DS factor}) can be checked by using the transformation 

\begin{equation}\label{gauge subst}
  \delta\epsilon_{n+1}^{\mu\nu} = p^{\mu}\Lambda^{\nu} + p^{\nu}\Lambda^{\mu} \qquad \text{or} \qquad \delta\epsilon_{n+2}^{\mu\nu} = q^{\mu}\Lambda^{\nu} + q^{\nu}\Lambda^{\mu}.
\end{equation}
 Unlike single soft case where momentum conservation is required to prove gauge invariance, no such consideration is needed for double soft case. The terms in 
 (\ref{gravity DS factor}) simply cancel among themselves after the substitution of (\ref{gauge subst}).

\end{itemize}
In the next section we investigate the above features in details by calculating the Feynman diagrams for linearized perturbative Einstein gravity. We will find that CHY expression (\ref{gravity DS factor}) of the double soft factor is actually the sub-leading order term at tree level. Unlike the leading order term, the sub-leading term does not appear as product of two single soft factors. Moreover this sub-leading term not only comes from two gravitons going soft at the same rate but also receives contribution from the processes where one graviton is taken to be soft at a faster rate than the other. Similar analyses also appear in \cite{Volovich:string, Klose:gluon} where particles are taken soft in succession.    

\section{Double Soft Limit from Feynman Diagrams for Gravity}
\label{sec:Feynman gravity}

Here we compute the double soft limit of gravity from the Feynman diagrams. The action for Einstein Hilbert gravity in four dimension is 

\begin{equation}\label{EH action}
 S_{EH} = \frac{2}{\kappa^{2}}\int\mathrm{d}^{4}x\sqrt{-g}R.
\end{equation}
The Feynman rules for linearized gravity (\ref{g pert}) are given in \cite{Sannan} where the conventions of \cite{DeWitt} are used. Every three-point
vertex is of $\mathcal{O}(\kappa) $ and four-point vertex is of $\mathcal{O}(\kappa^{2}) $. When looking for scattering amplitude in the double soft 
limit it is sufficient for our purpose to consider only three and four point vertices because they are the ones to contribute at the leading order. Given
$n$ external legs there can not be any higher than four-point vertex in each leg from where two soft gravitons can be emitted because in that case number of hard particles will exceed $n$.
In the perturbative linearized gravity in the double soft limit there are two parameters, coupling constant, $\kappa$ and energy scale of soft gravitons, 
$\tau$ and the dominating term, as we will see below, is of $\mathcal{O}\left(\frac{\kappa^{2}}{\tau}\right)$.

Following are the relevant Feynman diagrams (momenta at all external legs are outgoing):

\begin{itemize}

\item \textbf{4 point vertex}
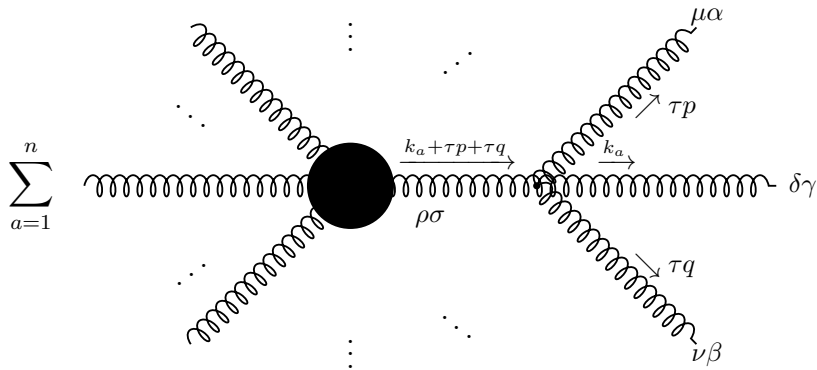
\begin{figure}[H]
 \begin{center}
  \begin{tikzpicture}[line width=0.7 pt, scale=0.7]
  \node at (-6,0) {$\mathlarger{\mathlarger{\sum\limits_{a=1}^{n}}} $};
   \filldraw[] (0,0) circle(0.8);
   \draw[graviton] (0,0) -- (1.5,0)node[below=0.2]{$\rho\sigma$} -- (2,0)node[above=.1]{$\xrightarrow{k_{a}+\tau p+\tau q}$}
		--(5,0)node[above=0.1]{$\xrightarrow{k_{a}}$} --(8,0);
   \node at (8.5,0) {$\delta\gamma$};
   \draw[graviton] (-5,0) -- (0,0);
   \draw[graviton] (-3,-3) -- (0,0);
   \draw[graviton] (-3,3) -- (0,0);
   \filldraw[] (3.5,0) circle(0.05);
   \draw[graviton] (3.5,0) -- (5,1.5)node[right=0.1] {$\nearrow\tau p$} -- (6.5,3);
   \draw[graviton] (3.5,0) -- (5,-1.5)node[right=0.1] {$\searrow\tau q$} -- (6.5,-3);
   \node at (6.7,3.2) {$\mu\alpha$};
   \node at (6.7,-3.2) {$\nu\beta$};
   \node at (-3,-1.5) {$\mathlarger{\mathlarger{\iddots}}$};
   \node at (-3,1.5) {$\mathlarger{\mathlarger{\ddots}}$};
   \node at (0,3) {$\mathlarger{\mathlarger{\vdots}}$};
   \node at (0,-3) {$\mathlarger{\mathlarger{\vdots}}$};
   \node at (2,-2.5) {$\mathlarger{\mathlarger{\ddots}}$};
   \node at (2,2.5) {$\mathlarger{\mathlarger{\iddots}}$};
  \end{tikzpicture}
 \end{center}
\end{figure}

\begin{equation}\label{scat 1}
 \approx -4\frac{\kappa^{2}}{\tau} \sum\limits_{a=1}^{n} \frac{\epsilon_{n+1}.\epsilon_{n+2} \: \epsilon_{n+1}.k_{a} \: \epsilon_{n+2}.k_{a}}{k_{a}.(p+q)} 
	    (M_{n})_{\rho\sigma}\epsilon_{a}^{\rho\sigma} + \kappa^{2} \mathcal{O}(1)
\end{equation}

\item \textbf{3 point vertex (I)}

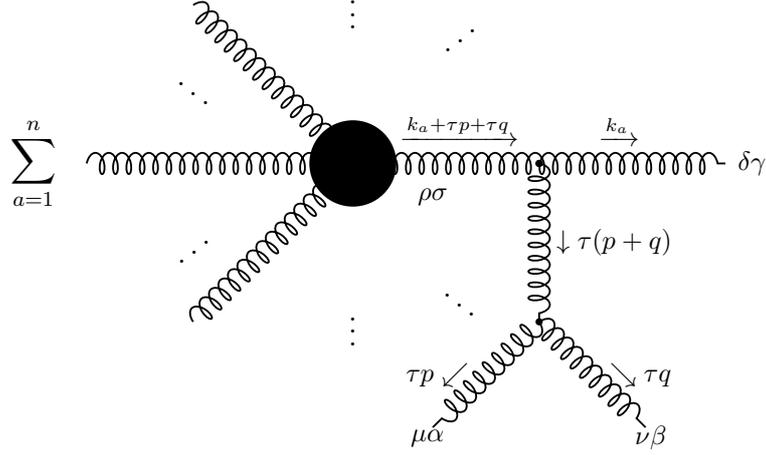
 \begin{figure}[H]
\begin{center}
\begin{tikzpicture}[line width=0.7 pt, scale=0.7]
\node at (-6,0) {$\mathlarger{\mathlarger{\sum\limits_{a=1}^{n}}} $};
   \filldraw[] (0,0) circle(0.8);
   \draw[graviton] (0,0) -- (1.5,0)node[below=0.2]{$\rho\sigma$} -- (2,0)node[above=.1]{$\xrightarrow{k_{a}+\tau p+\tau q}$}
		--(5,0)node[above=0.1]{$\xrightarrow{k_{a}}$} --(7,0);
   \node at (7.5,0) {$\delta\gamma$};
   \draw[graviton] (-5,0) -- (0,0);
   \draw[graviton] (-3,-3) -- (0,0);
   \draw[graviton] (-3,3) -- (0,0);
   \filldraw[] (3.5,0) circle(0.05);
\draw[graviton] (3.5,0) -- (3.5,-1.5)node[right=0.1]{$\downarrow \tau(p+q)$} -- (3.5,-3);
\filldraw[] (3.5,0) circle(0.05);
\draw[graviton] (3.5,-3) -- (2.5,-4)node[left=0.1]{$ \tau p\swarrow$} -- (1.5,-5);
\draw[graviton] (3.5,-3) -- (4.5,-4)node[right=0.1]{$ \searrow\tau q$} -- (5.5,-5);
\filldraw[] (3.5,-3) circle(0.05);
\node at (1.4,-5.2) {$\mu\alpha$};
\node at (5.6,-5.2) {$\nu\beta$};
\node at (-3,-1.5) {$\mathlarger{\mathlarger{\iddots}}$};
   \node at (-3,1.5) {$\mathlarger{\mathlarger{\ddots}}$};
   \node at (0,3) {$\mathlarger{\mathlarger{\vdots}}$};
   \node at (0,-3) {$\mathlarger{\mathlarger{\vdots}}$};
   \node at (2,-2.5) {$\mathlarger{\mathlarger{\ddots}}$};
   \node at (2,2.5) {$\mathlarger{\mathlarger{\iddots}}$};
\end{tikzpicture}
\end{center}
\end{figure}

\begin{eqnarray}\label{scat 2}
 & \approx & \frac{\kappa^{2}}{\tau} \sum\limits_{a=1}^{n} \frac{1}{p.q \: k_{a}.(p+q)} \left[ -(\epsilon_{n+1}.\epsilon_{n+2})^{2} k_{a}.p \: k_{a}.q 
		+ (\epsilon_{n+1}.q)^{2} (\epsilon_{n+2}.k_{a})^{2} + (\epsilon_{n+1}.k_{a})^{2} (\epsilon_{n+2}.p)^{2} \right. \nonumber \\
		&& \left. \phantom{\frac{\kappa^{2}}{\tau} \sum\limits_{a=1}^{n} \frac{1}{p.q \: k_{a}.(p+q)} \left[ \right.} 
		-2 \epsilon_{n+1}.\epsilon_{n+2} \biggl\{ \epsilon_{n+1}.k_{a} \: \epsilon_{n+2}.p \: k_{a}.p 
		+ \epsilon_{n+1}.q \: \epsilon_{n+2}.k_{a} \: k_{a}.q\biggr\} \right. \nonumber \\ 
		&& \left. \phantom{\frac{\kappa^{2}}{\tau} \sum\limits_{a=1}^{n} \frac{1}{p.q \: k_{a}.(p+q)} \left[ \right.} 
		-2 \: \epsilon_{n+1}.q \: \epsilon_{n+2}.p \: \epsilon_{n+1}.k_{a} \: \epsilon_{n+2}.k_{a} \right]
		(M_{n})_{\rho\sigma}\epsilon_{a}^{\rho\sigma} \nonumber\\
		&& + \: 2\frac{\kappa^{2}}{\tau}\sum\limits_{a=1}^{n} \frac{\epsilon_{n+1}.\epsilon_{n+2} \: \epsilon_{n+1}.k_{a} \: \epsilon_{n+2}.k_{a}}{k_{a}.(p+q)} 
		(M_{n})_{\rho\sigma}\epsilon_{a}^{\rho\sigma} + \kappa^{2} \mathcal{O}(1)
\end{eqnarray}

\item \textbf{3 point vertex (II)}

 \begin{figure}[H]
 \begin{center}
  \begin{tikzpicture}[line width=0.7 pt, scale=0.7]
  \node at (-6,0) {$\mathlarger{\mathlarger{\sum\limits_{a=1}^{n}}} $};
   \filldraw[] (0,0) circle(0.8);
   \draw[graviton] (0,0) -- (1.5,0)node[below=0.2]{$\rho\sigma$} -- (2,0)node[above=.1]{$\xrightarrow{k_{a}+\tau p+\tau q}$}
		-- (5,0)node[above=0.1]{$ \xrightarrow{k_{a}+\tau q}$}  --(8,0)node[above=0.1]{$\xrightarrow{k_{a}}$} --(9,0);
   \node at (9.5,0) {$\delta\gamma$};
   \draw[graviton] (-5,0) -- (0,0);
   \draw[graviton] (-3,-3) -- (0,0);
   \draw[graviton] (-3,3) -- (0,0);
   \node at (6.7,3.7) {$\mu\alpha$};
   \node at (8.7,-3.7) {$\nu\beta$};
   \draw[graviton] (3.5,0) -- (5,1.5)node[right=0.1] {$\nearrow\tau p$} -- (6.5,3);
   \draw[graviton] (5.5,0) -- (7,-1.5)node[right=0.1] {$\searrow\tau q$} -- (8.5,-3);
   \filldraw[] (3.5,0) circle(0.05);
   \filldraw[] (5.5,0) circle(0.05);
   \node at (12,0) { $ + $ (sym $ p \leftrightarrow q $)};
   \node at (-3,-1.5) {$\mathlarger{\mathlarger{\iddots}}$};
   \node at (-3,1.5) {$\mathlarger{\mathlarger{\ddots}}$};
   \node at (0,3) {$\mathlarger{\mathlarger{\vdots}}$};
   \node at (0,-3) {$\mathlarger{\mathlarger{\vdots}}$};
   \node at (2,-2.5) {$\mathlarger{\mathlarger{\ddots}}$};
   \node at (2,2.5) {$\mathlarger{\mathlarger{\iddots}}$};
   \end{tikzpicture}
  \end{center}
  \end{figure}
  
  \begin{eqnarray}\label{scat 3}
   & \approx & \frac{\kappa^{2}}{\tau}\sum\limits_{a=1}^{n} \frac{1}{k_{a}.(p+q)} \left[ \frac{2 \: \epsilon_{n+1}.q \: \epsilon_{n+1}.k_{a} (\epsilon_{n+2}.k_{a})^{2}}
			    {k_{a}.q}  + \frac{2 \: (\epsilon_{n+1}.k_{a})^{2} \: \epsilon_{n+2}.p \: \epsilon_{n+2}.k_{a}}{k_{a}.p} \right. \nonumber\\
			    && \left. \phantom{\frac{\kappa^{2}}{\tau}\sum\limits_{a=1}^{n} \frac{1}{k_{a}.(p+q)} \left[ \right.}
			    -\frac{(\epsilon_{n+1}.k_{a})^{2} (\epsilon_{n+2}.k_{a})^{2} p.q}{k_{a}.p \: k_{a}.q} \right]
			    (M_{n})_{\rho\sigma}\epsilon_{a}^{\rho\sigma} + \kappa^{2} \mathcal{O}(1)		    
  \end{eqnarray}
  
  The last term comes from the expansion of the propagator in the denominator 
  \begin{equation}
   \frac{1}{(k_{a}+ \tau p + \tau q)^{2}} \approx \frac{1}{\tau \: k_{a}.(p+q)}\left[ 1 - \frac{\tau \: p.q}{k_{a}.(p+q)} + \mathcal{O}(\tau^{2}) \right]. 
  \end{equation}
  
  \paragraph{Comments:}
  There is a subtlety with the above diagram. Strictly speaking the leading order term corresponding to this diagram is of $\mathcal{O}\left(\frac{\kappa^{2}}{\tau^{2}}\right)$,
  given by
  
  \begin{equation}\label{anomalous}
   \frac{\kappa^{2}}{\tau^{2}} \sum\limits_{a=1}^{n} \frac{(\epsilon_{n+1}.k_{a})^{2}(\epsilon_{n+2}.k_{a})^{2}}{k_{a}.p \: k_{a}.q}
				(M_{n})_{\rho\sigma}\epsilon_{a}^{\rho\sigma},
  \end{equation}
which is product of two single soft factors where the soft gravitons are emitted from the same external leg. It is interesting to note that
this term by itself is not gauge invariant, hence can not occur alone in the scattering amplitude. So to preserve gauge invariance we need to 
add terms coming from the following Feynman diagram:

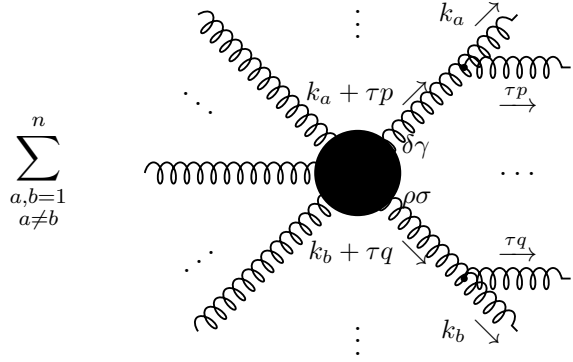
\begin{figure}[H]
 \begin{center}
  \begin{tikzpicture}[line width=0.7 pt, scale=0.7 pt]
  \node at (-6,0) {$\mathlarger{\mathlarger{\sum\limits_{\substack{a,b=1\\a \ne b}}^{n}}} $};
   \filldraw[] (0,0) circle (0.8);
   \draw[graviton] (-3,3) -- (0,0);
   \draw[graviton] (-4,0) -- (0,0);
   \draw[graviton] (-3,-3) -- (0,0);
   \draw[graviton] (0,0) --(0.5,0.5)node[right=0.1]{$\delta\gamma$} -- (1.5,1.5)node[left=0.001] {$k_{a}+\tau p\nearrow$}-- (3,3)node[left=0.1] {$k_{a}\nearrow$} --(3,3);
   \draw[graviton] (0,0) --(0.5,-0.5)node[right=0.1]{$\rho\sigma$}-- (1.5,-1.5)node[left=0.001] {$k_{b}+\tau q\searrow$}-- (3,-3)node[left=0.1] {$k_{b}\searrow$} -- (3,-3);
   \draw[graviton] (2,2) --(3,2)node[below=0.1]{$\xrightarrow{\tau p}$}-- (4,2);
   \draw[graviton] (2,-2) --(3,-2)node[above=0.1]{$\xrightarrow{\tau q}$}-- (4,-2);
   \filldraw[] (2,2) circle(0.05);
   \filldraw[] (2,-2) circle(0.05);
   \node at (-3,-1.5) {$\mathlarger{\mathlarger{\iddots}}$};
   \node at (-3,1.5) {$\mathlarger{\mathlarger{\ddots}}$};
   \node at (0,3) {$\mathlarger{\mathlarger{\vdots}}$};
   \node at (0,-3) {$\mathlarger{\mathlarger{\vdots}}$};
   \node at (3,0) {$\mathlarger{\mathlarger{\ldots}}$};
  \end{tikzpicture}
 \end{center}
\end{figure}
This diagram together with expression (\ref{anomalous}) gives

\begin{equation}\label{leading ds}
 \frac{\kappa^{2}}{\tau^{2}} \sum\limits_{\substack{a,b=1\\a\ne b}}^{n} \frac{(\epsilon_{n+1}.k_{a})^{2}(\epsilon_{n+2}.k_{b})^{2}}{k_{a}.p \: k_{b}.q}	M_{n}
\end{equation}
which is precisely the product of two single soft factors and therefore gauge invariant. 

Interestingly this scattering is not a \textit{local} process, as mentioned towards the end of sec(\ref{sec:EH}), in the sense that two soft gravitons are coming 
from different external legs. Therefore the term (\ref{anomalous}) is actually a part of this \textit{nonlocal} process. Due to this reason we do not find this 
term appearing in the CHY expression (\ref{gravity DS factor}). 
  
  \item \textbf{3 point vertex (III)}
  
  After summing over all the above diagrams we can account for almost all the terms of CHY expression in Eq.(\ref{gravity DS factor}) except one, which is of the 
  form $ \sim \frac{1}{\tau\: p.q}\biggl\{ \frac{(\epsilon_{n+1}.q)^{2}\:(\epsilon_{n+2}.p)^{2}}{k_{a}.q} + \frac{(\epsilon_{n+1}.k_{a})^{2}\: (\epsilon_{n+2}.p)^{2}}{k_{a}.p} \biggr\}$.
  Also this term is required for preserving gauge invariance. This motivates us to look for the following diagram\footnote{We thank Alok Laddha for his suggestion to look into this process.}:  
  
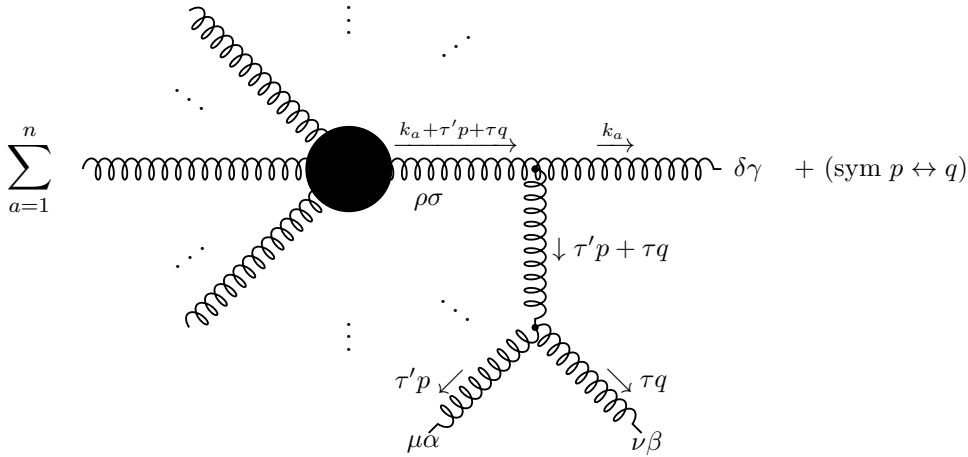
  \begin{figure}[H]
\begin{center}
\begin{tikzpicture}[line width=0.7 pt, scale=0.7]
\node at (-6,0) {$\mathlarger{\mathlarger{\sum\limits_{a=1}^{n}}} $};
   \filldraw[] (0,0) circle(0.8);
   \draw[graviton] (0,0) -- (1.5,0)node[below=0.2]{$\rho\sigma$} -- (2,0)node[above=.1]{$\xrightarrow{k_{a}+\tau' p+\tau q}$}
		--(5,0)node[above=0.1]{$\xrightarrow{k_{a}}$} --(7,0);
   \node at (7.5,0) {$\delta\gamma$};
   \draw[graviton] (-5,0) -- (0,0);
   \draw[graviton] (-3,-3) -- (0,0);
   \draw[graviton] (-3,3) -- (0,0);
   \filldraw[] (3.5,0) circle(0.05);
\draw[graviton] (3.5,0) -- (3.5,-1.5)node[right=0.1]{$\downarrow \tau'p+\tau q$} -- (3.5,-3);
\filldraw[] (3.5,0) circle(0.05);
\draw[graviton] (3.5,-3) -- (2.5,-4)node[left=0.1]{$ \tau' p\swarrow$} -- (1.5,-5);
\draw[graviton] (3.5,-3) -- (4.5,-4)node[right=0.1]{$ \searrow\tau q$} -- (5.5,-5);
\filldraw[] (3.5,-3) circle(0.05);
\node at (1.4,-5.2) {$\mu\alpha$};
\node at (5.6,-5.2) {$\nu\beta$};
\node at (10,0) { $ + $ (sym $ p \leftrightarrow q $)};
   \node at (-3,-1.5) {$\mathlarger{\mathlarger{\iddots}}$};
   \node at (-3,1.5) {$\mathlarger{\mathlarger{\ddots}}$};
   \node at (0,3) {$\mathlarger{\mathlarger{\vdots}}$};
   \node at (0,-3) {$\mathlarger{\mathlarger{\vdots}}$};
   \node at (2,-2.5) {$\mathlarger{\mathlarger{\ddots}}$};
   \node at (2,2.5) {$\mathlarger{\mathlarger{\iddots}}$};
\end{tikzpicture}
\end{center}
\end{figure}

These describe the process where one of the soft gravitons is more softer than the other and the softer one is being emitted from the relatively harder one.
Let the momentum of the $ (n+1) $th particle goes as $ \tau' p $ and that of $ (n+2) $th goes as $ \tau q $, where $ \tau' > \tau $ in the limit 
both $ \tau' $, $ \tau \rightarrow 0 $. Then the overall scaling of the amplitude will go as $ \frac{1}{\tau} $. Upon symmetrizing between $ p $ and $ q $
we get  

\begin{equation}\label{scat 4}
 -\frac{\kappa^{2}}{\tau \: p.q}\sum\limits_{a=1}^{n}\biggl\{ \frac{(\epsilon_{n+1}.q)^{2}\:(\epsilon_{n+2}.p)^{2}}{k_{a}.q} + \frac{(\epsilon_{n+1}.k_{a})^{2}\:
			    (\epsilon_{n+2}.p)^{2}}{k_{a}.p} \biggr\}(M_{n})_{\rho\sigma}\epsilon_{a}^{\rho\sigma} + \kappa^{2} \mathcal{O}(1)
\end{equation}
which is precisely the term we are looking for.
\end{itemize}

Adding together the expressions (\ref{scat 1}), (\ref{scat 2}), (\ref{scat 3}) and (\ref{scat 4})  we recover the CHY expression for the double soft factor (\ref{gravity DS factor})
exactly. This matching of the results is significant in the sense that it helps to clarify the correspondence between the gravity integrand (\ref{gravity integrand}) of 
CHY formula and the Einstein Hilbert action (\ref{EH action}). Moreover we emphasize on the fact that in pure gravity due to the presence of $ggg$ vertices we see \textit{nonlocal}
processes as described earlier, because of which CHY graviton scattering amplitude in the double soft limit corresponds to a subset of possible Feynman diagrams. Unlike 
gravity this peculiarity is not encountered for soft photon emissions in Einstein Maxwell theory where two soft photons are always emitted from single external leg through $AAg$ vertices. 

\section{Discussion and Conclusion}
\label{discussion}

Cachazo et al have pioneered an innovative method for calculating tree level scattering amplitude, for a wide variety of theories including gravity, without pertaining to 
the explicit computation of Feynman diagrams which grow enormously large in number and are cumbersome for higher point amplitudes. A remarkable aspect of CHY formalism
is one can get the soft limits of these scattering amplitudes much more conveniently than can be done from Feynman diagrammatic. The prescriptions for 
taking single and double soft limits have been given by Cachazo and his collaborators. In this paper we have applied their prescription to derive the double
soft limit of the gravity scattering amplitude. Like the case of single soft limit, here too, 
the result we get is of $\mathcal{O}\left(\frac{1}{\tau}\right)$ and not $\mathcal{O}\left(\frac{1}{\tau^{2}}\right)$ as one might have guessed. 
Also, CHY expression for double soft limit implies the two soft gravitons should necessarily come from single external graviton leg. In case of gravity 
there are scattering processes where soft gravitons are emitted from different external legs, and these are not included in CHY formula.

Computing the relevant Feynman diagrams from linearized perturbative Einstein gravity some interesting features stand out:

\begin{itemize}
 \item There are some diagrams which contribute at $\mathcal{O}\left(\frac{1}{\tau^{2}}\right)$. But these correspond to the processes where soft 
 gravitons are coming from different external legs and hence CHY expression does not contain these terms. 
 
 \item At $\mathcal{O}\left(\frac{1}{\tau}\right)$ we found a particular process where the scaling of energies of the soft gravitons are different; one particle 
 goes softer than the other one. This diagram is necessary to make the scattering amplitude gauge invariant and also turns out that this term is included in 
 the CHY answer. 
 
\end{itemize}

Finally the fact that considering Feynman diagrams coming from Einstein Hilbert action, the double soft factor of the scattering amplitude matches precisely with that
of CHY answer helps us to clarify the correspondence between the gravity integrand proposed by Cachazo et al and the Einstein gravity. 

Loop corrections to single soft theorems have been studied for gluon and gravity amplitudes in \cite{Bern:loopcor, He:loopcor, Bianchi:loops}. From their analyses it is evident that leading soft theorems are protected from loop corrections but sub-leading ones require corrections at loop level. One-loop corrections in CHY formalism have been studied in \cite{He:loop, CHY:loop} for bi-adjoint scalars and gluon amplitudes. It will be interesting to study how loop corrections can be incorporated in the present context and this will be helpful to learn about the universality of double soft theorem presented here. 

In \cite{Strominger:Scat, Strominger:ST, Strominger:2014pwa, Campiglia:2015yka, Campiglia:2014yka, Campiglia:2016jdj, Campiglia:2015kxa} the equivalence between 
Weinberg's soft graviton theorem  and BMS supertranslation Ward identity of S matrix at asymptotic null infinity is established. It will be extremely interesting 
to see if the double soft theorems are related to Ward identities associated to certain symmetries. As double soft theorems are \enquote{non-local} function in 
conformal $\mathbb{S}^{2}$ at null infinity, such symmetries can perhaps leave a very intricate structure distinct from BMS.

\section{Acknowledgements}

I am immensely grateful to Alok Laddha for suggesting me this problem, for his
constant supervision and guidance through out the entire course of this project and for his vast contribution to this paper.
I am also grateful to A. P. Balachandran, S. Kalyana Rama and Balachandran Sathiapalan for various helpful discussions 
and valuable suggestions. I would also like to thank my friends especially Taushif Ahmed, Pulak Banerjee, Atanu Bhatta, Prasanna Kumar Dhani, Sanjoy Mandal and 
Narayan Rana for useful discussions regarding Feynman diagrams. Finally I would like to thank the anonymous referee for providing valuable suggestions.

\appendix
\section{Appendix}

\subsection{Derivation of Eq.(\ref{grav ds Pf expansion})}
\label{sec:appendix gravity integrand}

At leading order in $ \tau $ the structure of matrix $ \Psi_{n+2} $ is \cite{Cachazo:supp}

\begin{equation}\label{finite rho matrix}
 \Psi_{n+2} \approx
 \left(
 \begin{array}{c:c:c|c:c:c}
  (A_n)_{ab} & \frac{\tau\: k_{a}.p}{\sigma_{a} - \rho} & \frac{\tau\: k_{a}.q}{\sigma_{a} - \rho} & (-C_{n}^{T})_{ab} & \frac{-\epsilon_{n+1}.k_{b}}{\rho-\sigma_{b}} & \frac{-\epsilon_{n+2}.k_{b}}{\rho-\sigma_{b}} \\
  \hdashline\\
  \frac{\tau \: p.k_{b}}{\rho - \sigma_{b}} & 0 & \frac{-\tau \: p.q}{\xi_{1}} & \frac{-\tau \: \epsilon_{a}.p}{\sigma_{a}-\rho} & -C_{n+1,n+1} & \frac{-\epsilon_{n+2}.p}{\xi_{1}}\\
  \hdashline\\
  \frac{\tau \: q.k_{b}}{\rho - \sigma_{b}} & \frac{\tau \: p.q}{\xi_{1}} & 0 & \frac{-\tau \: \epsilon_{a}.q}{\sigma_{a}-\rho} & \frac{\epsilon_{n+1}.q}{\xi_{1}} & -C_{n+2,n+2} \\
  \hline\\
  (C_{n})_{ab} & \frac{\tau \: \epsilon_{a}.p}{\sigma_{a}-\rho} & \frac{\tau \: \epsilon_{a}.q}{\sigma_{a}-\rho} & (B_{n})_{ab} & \frac{\epsilon_{a}.\epsilon_{n+1}}{\sigma_{a}-\rho} & \frac{\epsilon_{a}.\epsilon_{n+2}}{\sigma_{a}-\rho} \\
  \hdashline\\
  \frac{\epsilon_{n+1}.k_{b}}{\rho-\sigma_{b}} & C_{n+1,n+1} & \frac{-\epsilon_{n+1}.q}{\xi_{1}} & \frac{\epsilon_{n+1}.\epsilon_{b}}{\rho-\sigma_{b}} & 0 & \frac{\epsilon_{n+1}.\epsilon_{n+2}}{-\tau\:\xi_{1}} \\
  \hdashline\\
  \frac{\epsilon_{n+2}.k_{b}}{\rho-\sigma_{b}} & \frac{\epsilon_{n+2}.p}{\xi_{1}} & C_{n+2,n+2} & \frac{\epsilon_{n+2}.\epsilon_{b}}{\rho-\sigma_{b}} & \frac{\epsilon_{n+1}.\epsilon_{n+2}}{\tau\:\xi_{1}} & 0 
 \end{array}
 \right)
\end{equation}
Now we use the identity for Pfaffian on any $ 2m \times 2m $ matrix $ E $

\begin{equation}\label{Pf expansion}
 \mathrm{Pf}(E) = \sum\limits_{q=1}^{2m} (-1)^{q} e_{pq} \mathrm{Pf}(E^{pq}_{pq}),
\end{equation}
where $e_{pq}$ is the element of the matrix $E$ at the $p$th row and $q$th column.
First we make an expansion of the Pfaffian of $ \Psi_{n+2} $ along the $ (n+2) $th row to get upto leading order in $ \tau $

\begin{equation}
 \mathrm{Pf'}\Psi_{n+2} = \frac{\tau\: p.q}{\xi_{1}}\mathrm{Pf'}(\Psi_{n+2})^{n+2,n+1}_{n+2,n+1} - \frac{\epsilon_{n+1}.q}{\xi_{1}}\mathrm{Pf'}(\Psi_{n+2})^{n+2,n+3}_{n+2,n+3}
			  -C_{n+2,n+2}\mathrm{Pf'}(\Psi_{n+2})^{n+2,2n+4}_{n+2,2n+4}.
\end{equation}
Again each of the reduced Pfaffians can be further expanded as

\begin{eqnarray}
 \mathrm{Pf'}(\Psi_{n+2})^{n+2,n+1}_{n+2,n+1} & = & -\frac{\epsilon_{n+1}.\epsilon_{n+2}}{\tau \: \xi_{1}} \mathrm{Pf'}\Psi_{n} + \mathcal{O}(1) \nonumber\\
 \mathrm{Pf'}(\Psi_{n+2})^{n+2,2n+3}_{n+2,2n+3} & = & -\frac{\epsilon_{n+2}.p}{\xi_{1}}\mathrm{Pf'}\Psi_{n} + \mathcal{O}(\tau) \nonumber\\
 \mathrm{Pf'}(\Psi_{n+2})^{n+2,2n+4}_{n+2,2n+4} & = & -C_{n+1,n+1}\mathrm{Pf'}\Psi_{n} + \mathcal{O}(\tau).
\end{eqnarray}
The two diagonal terms of the matrix $ C_{n+2} $ can be approximated as

\begin{eqnarray}
 C_{n+1,n+1} & = & -\sum\limits_{i=1}^{n} \frac{\epsilon_{n+1}.p_{i}^{\bot}}{\rho - \sigma_{b}} \nonumber\\
 C_{n+2,n+2} & = & -\sum\limits_{i=1}^{n} \frac{\epsilon_{n+2}.q_{i}^{\bot}}{\rho - \sigma_{b}}
\end{eqnarray}
where $ p_{i}^{\bot} = k_{i} - \frac{p.k_{i}}{p.q}q $ and $ q_{i}^{\bot} = k_{i} - \frac{q.k_{i}}{p.q}p $. Putting together all these expressions in Eq.(\ref{Pf expansion}) we get Eq.(\ref{grav ds Pf expansion})

\subsection{Derivation of Double Soft Factor}
\label{sec: app DS factor}

Using Eq.(\ref{double soft factorization}) and Eq.(\ref{grav ds Pf expansion}) we get 

\begin{equation}
 S^{*(0)} = -\frac{1}{\tau}\oint \frac{\mathrm{d\rho}}{2\pi i} \frac{\xi_{1}^{2}}{p.q} \: \sum\limits_{a=1}^{n}\frac{k_{a}.(p+q)}{\rho-\sigma_{a}} \left[ \frac{\epsilon_{n+1}.q \:
	      \epsilon_{n+2}.p - \epsilon_{n+1}.\epsilon_{n+2} \: p.q}{\xi_{1}^{2}} + \sum\limits_{i,j =1}^{n} \frac{\epsilon_{n+1}.p_{i}^{\bot} \:
		\epsilon_{n+2}.q_{j}^{\bot}}{(\rho-\sigma_{i})(\rho-\sigma_{j})} \right]^{2}
\end{equation}
There are no contributions from higher order poles in the contour integral, only simple poles at $ \rho = \sigma_{a} $ contribute. Substituting the solutions of $ \xi_{1} $ from
Eq.(\ref{xi sol}) and using their product for $ \xi_{1}^{2} $ we get

\begin{eqnarray}
 S^{*(0)} & = & -\frac{1}{\tau} \sum\limits_{a=1}^{n}\frac{1}{k_{a}.(p+q)}\left[ - \frac{\left( \epsilon_{n+1}.q \: \epsilon_{n+2}.p - \epsilon_{n+1}.\epsilon_{n+2} \: p.q\right)^{2} k_{a}.p \: k_{a}.q}{(p.q)^{3}} \right. \nonumber\\
	      && \left. \phantom{-\frac{1}{\tau} \sum\limits_{a=1}^{n}\frac{1}{k_{a}.(p+q)}\left[ \right.} +2 \frac{\left(\epsilon_{n+1}.q \: \epsilon_{n+2}.p - \epsilon_{n+1}.\epsilon_{n+2} \: p.q\right)\epsilon_{n+1}.p_{a}^{\bot} \: \epsilon_{n+2}.q_{a}^{\bot}}{p.q} \right. \nonumber\\
	      && \left. \phantom{-\frac{1}{\tau} \sum\limits_{a=1}^{n}\frac{1}{k_{a}.(p+q)}\left[ \right.}-\frac{\left(\epsilon_{n+1}.p_{a}^{\bot} \: \epsilon_{n+2}.q_{a}^{\bot}\right)^{2}p.q}{k_{a}.p \: k_{a}.q} \right]
\end{eqnarray}
After simplification the above expression reduces to Eq.(\ref{gravity DS factor}).

\subsection{Soft Factor from Pole at Infinity}
\label{sec:app pole inf}

In case of gravity there exists a simple pole at $ \rho = \infty $. The equation $ f_{n+1} - f_{n+2} =0 $ leads to $ \xi = \pm 2 i \rho + \mathcal{O}(1) $.
Then the Pfaffian of $ \Psi_{n+2} $ can be expanded as

\begin{equation}
 \mathrm{Pf'}(\Psi_{n+2}) = \frac{\tau^{2} \: p.q \: \epsilon_{n+1}.\epsilon_{n+2}}{4 \rho^{2}} \mathrm{Pf'}(\Psi_{n}) + \mathcal{O}\left(\frac{1}{\rho^{4}}\right).
\end{equation}
Then from Eq.(\ref{Pole inf}) we get

\begin{equation}
 (M_{n+2})_{\infty} = -\frac{1}{3} (\epsilon_{n+1}.\epsilon_{n+2})^{2} M_{n}
\end{equation}
which is clearly subleading in order $ \tau $ as compared to Eq.(\ref{gravity DS factor}).

\subsection{Feynman Rules for EM}
\label{sec:EM Feynman}

The Feynman rules for Yang-Mills theory coupled to gravity have been derived in \cite{YM}. In the same way Feynman rules for Einstein Maxwell 
theory coupled to gravity can be derived. 

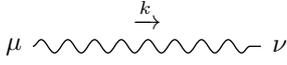
\begin{figure}[H]
   \begin{tikzpicture}[line width=0.5 pt, scale=0.5]
    \draw[vector] (-3,0) -- (0,0)node[above=0.1]{$ \xrightarrow{k} $} -- (3,0);
    \node at (-3.5,0) {$ \mu $};
    \node at (3.5,0) {$ \nu $};
  \node at (20,0)
   { $ -\frac{i \:\eta_{\mu\nu}}{k^{2} + i\epsilon} $};
  \end{tikzpicture}
\end{figure}

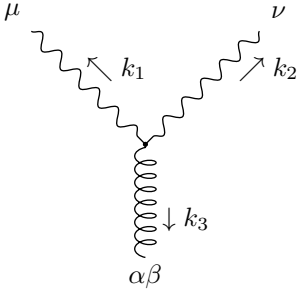
\begin{figure}[H]
 \begin{center}
  \begin{tikzpicture}[line width=0.5 pt, scale=0.5]
   \draw[vector] (-3,3) -- (-2,2)node[right=0.1]{$ \nwarrow k_{1} $}-- (0,0);
   \draw[vector] (3,3) -- (2,2)node[right=0.1]{$ \nearrow k_{2} $} -- (0,0);
   \draw[graviton] (0,-3) -- (0,-2)node[right=0.1]{$\downarrow  k_{3} $} -- (0,0);
   \filldraw[] (0,0) circle(0.05);
   \node at (-3.5,3.5) {$ \mu $};
   \node at (3.5,3.5) {$ \nu $};
   \node at (0,-3.5) {$ \alpha\beta $};
   \node at (20,0)
   {
   $\begin{aligned}
    -i\:\kappa & \left[ (\eta_{\mu\alpha}\eta_{\nu\beta} + \eta_{\mu\beta}\eta_{\nu\alpha} - \eta_{\mu\nu}\eta_{\alpha\beta})k_{1}.k_{2}
		      +\eta_{\mu\nu}(k_{1\alpha}k_{2\beta} + k_{1\beta}k_{2\alpha}) \right. \\
		& \left.\phantom{\left[ \right.} -k_{1\nu}k_{2\mu}\eta_{\alpha\beta} - k_{1\nu}(\eta_{\mu\alpha}k_{2\beta} + \eta_{\mu\beta}k_{2\alpha}) 
		    -k_{2\mu}(\eta_{\nu\alpha}k_{1\beta} + \eta_{\nu\beta}k_{1\alpha}) \right] 
    \end{aligned}$
    };
  \end{tikzpicture}
 \end{center}
\end{figure}

\subsection{Feynman Rules for Gravity}
\label{sec:grav feynman}

The Feynman rules for three and four point vertices are given in \cite{Sannan}.
Using Eq.(\ref{g pert}) we can write

\begin{equation}
 g^{\mu\nu} = \eta^{\mu\nu} - \kappa\: h^{\mu\nu} + \kappa^{2}h^{\mu\alpha}h_{\alpha}^{\phantom{\alpha}\nu} + \mathcal{O}(\kappa^{3})
\end{equation}
and

\begin{equation}
 \sqrt{-g} = 1 + \frac{\kappa}{2}h + \frac{\kappa^{2}}{8}\left(h^{2} - 2\: h^{\alpha\beta}h_{\alpha\beta}\right) + \mathcal{O}(\kappa^{3})
\end{equation}
where $ h = h^{\alpha}_{\phantom{\alpha}\alpha} $. The Ricci scalar can be expanded as

\begin{eqnarray}
 R & = & \kappa\left(\Box h - \partial_{\mu}\partial_{\nu}h^{\mu\nu} \right) \nonumber\\
	&& \phantom{\kappa\left(\Box h -\right.} + \kappa^{2} \left(\frac{1}{4}\partial_{\mu}h\partial^{\mu}h - \partial_{\mu}h^{\mu\nu}\partial_{\nu}h + 
	      \partial_{\mu}h^{\mu\nu}\partial^{\rho}h_{\nu\rho} - \frac{3}{4}\partial_{\mu}h_{\nu\rho}\partial^{\mu}h^{\nu\rho}\right. \nonumber\\
	&&   \left. \phantom{\phantom{\kappa\left(\Box h -\right.} \kappa^{2} \left(\frac{1}{4}\partial_{\mu}h\partial^{\nu}h - \right.} + 
	  \frac{1}{2}\partial_{\mu}h_{\nu\rho}\partial^{\nu}h^{\mu\rho} + 2\:h^{\mu\nu}\partial_{\mu}\partial^{\rho}h_{\rho\nu} - h^{\mu\nu}\Box h_{\mu\nu}\right) + \mathcal{O}(\kappa^{3}).
\end{eqnarray}
We work in harmonic (de Donder) gauge, where

\begin{equation}
 h^{\alpha}_{\phantom{\alpha}\mu, \alpha} - \frac{1}{2}h_{,\mu} = 0.
\end{equation}
Then the resulting graviton propagator, three and four point vertices are:

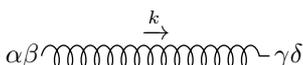
\begin{figure}[H]
   \begin{tikzpicture}[line width=0.5 pt, scale=0.5]
    \draw[graviton] (-3,0) -- (0,0)node[above=0.1]{$ \xrightarrow{k} $} -- (3,0);
    \node at (-3.5,0) {$ \alpha\beta $};
    \node at (3.5,0) {$ \gamma\delta $};
  \node at (20,0)
   { $ -\frac{i}{2} \frac{\eta_{\alpha\gamma} \eta_{\beta\delta} + \eta_{\alpha\delta}\eta_{\beta\gamma} - \eta_{\alpha\beta}\eta_{\gamma\delta}}
		    {k^{2} - i \epsilon} $};
  \end{tikzpicture}
\end{figure}

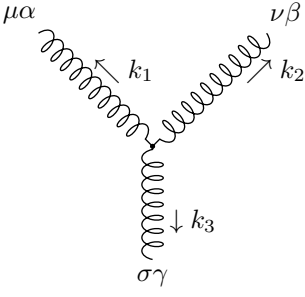
\begin{figure}[H]
 \begin{center}
  \begin{tikzpicture}[line width=0.5 pt, scale=0.5]
   \draw[graviton] (-3,3) -- (-2,2)node[right=0.1]{$ \nwarrow k_{1} $}-- (0,0);
   \draw[graviton] (3,3) -- (2,2)node[right=0.1]{$ \nearrow k_{2} $} -- (0,0);
   \draw[graviton] (0,-3) -- (0,-2)node[right=0.1]{$\downarrow  k_{3} $} -- (0,0);
   \filldraw[] (0,0) circle(0.05);
   \node at (-3.5,3.5) {$ \mu\alpha $};
   \node at (3.5,3.5) {$ \nu\beta $};
   \node at (0,-3.5) {$ \sigma\gamma $};
   \node at (20,0)
   {
   $\begin{aligned} \text{sym}
     & \left[ -\frac{1}{2}P_{3}(k_{1}.k_{2} \: \eta_{\mu\alpha} \eta_{\nu\beta} \eta_{\sigma\gamma}) -\frac{1}{2}P_{6}(k_{1\nu}k_{1\beta}\eta_{\mu\alpha}\eta_{\sigma\gamma})
		+\frac{1}{2}P_{3}(k_{1}.k_{2} \: \eta_{\mu\nu}\eta_{\alpha\beta}\eta_{\sigma\gamma}) \right.\\
     & \left.\phantom{\left[ \right.} + P_{6}(k_{1}.k_{2} \: \eta_{\mu\alpha}\eta_{\nu\sigma}\eta_{\beta\gamma}) + 2P_{3}(k_{1\nu}k_{1\gamma}\eta_{\mu\alpha}\eta_{\beta\sigma})
		-P_{3}(k_{1\beta}k_{2\mu}\eta_{\alpha\nu}\eta_{\sigma\gamma}) \right. \\
      & \left.\phantom{\left[ \right.} + P_{3}(k_{1\sigma}k_{2\gamma}\eta_{\mu\nu}\eta_{\alpha\beta}) + P_{6}(k_{1\sigma}k_{1\gamma}\eta_{\mu\nu}\eta_{\alpha\beta}) 
		+ 2P_{6}(k_{1\nu}k_{2\gamma}\eta_{\beta\mu}\eta_{\alpha\sigma}) \right. \\
      & \left. \phantom{\left[ \right.} + 2P_{3}(k_{1\nu}k_{2\mu}\eta_{\beta\sigma}\eta_{\gamma\alpha}) - 2P_{3}(k_{1}.k_{2} \: \eta_{\alpha\nu}\eta_{\beta\sigma}\eta_{\gamma\mu}) \right]
    \end{aligned} $
    };
  \end{tikzpicture}
 \end{center}
\end{figure}

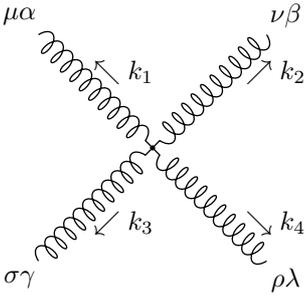
\begin{figure}[H]
 \begin{center}
  \begin{tikzpicture}[line width=0.5pt, scale=0.5]
   \draw[graviton] (-3,3) -- (-2,2)node[right=0.1]{$ \nwarrow k_{1} $}-- (0,0);
   \draw[graviton] (3,3) -- (2,2)node[right=0.1]{$ \nearrow k_{2} $} -- (0,0);
   \draw[graviton] (-3,-3) -- (-2,-2)node[right=0.1]{$\swarrow  k_{3} $} -- (0,0);
   \draw[graviton] (3,-3) -- (2,-2)node[right=0.1]{$\searrow  k_{4} $} -- (0,0);
   \filldraw[] (0,0) circle(0.05);
   \node at (-3.5,3.5) {$ \mu\alpha $};
   \node at (3.5,3.5) {$ \nu\beta $};
   \node at (-3.5,-3.5) {$ \sigma\gamma $};
   \node at (3.5,-3.5) {$ \rho\lambda $};
   \node at (20,0)
   {
   $ \begin{aligned}
       \text{sym} & \left[ -\frac{1}{4} P_{6}(k_{1}.k_{2} \: \eta_{\mu\alpha}\eta_{\nu\beta}\eta_{\sigma\gamma}\eta_{\rho\lambda}) 
		-\frac{1}{4}P_{12}(k_{1\nu}k_{1\beta}\eta_{\mu\alpha}\eta_{\sigma\gamma}\eta_{\rho\lambda}) 
		  -\frac{1}{2}P_{6}(k_{1\nu}k_{2\mu}\eta_{\alpha\beta}\eta_{\sigma\gamma}\eta_{\rho\lambda}) \right.\\
		  & \left. \phantom{\left[ \right.} + \frac{1}{4}P_{6}(k_{1}.k_{2} \: \eta_{\mu\nu}\eta_{\alpha\beta}\eta_{\sigma\gamma}\eta_{\rho\lambda})
		  + \frac{1}{2}P_{6}(k_{1}.k_{2}\: \eta_{\mu\alpha}\eta_{\nu\beta}\eta_{\sigma\rho}\eta_{\gamma\lambda})
		  + \frac{1}{2}P_{12}(k_{1\nu}k_{2\beta}\eta_{\mu\alpha}\eta_{\sigma\rho}\eta_{\gamma\lambda}) \right. \\
		  & \left.\phantom{\left[ \right.} + P_{6}(k_{1\nu}k_{2\mu}\eta_{\alpha\beta}\eta_{\sigma\rho}\eta_{\gamma\lambda})
		  -\frac{1}{2}P_{6}(k_{1}.k_{2}\: \eta_{\mu\nu}\eta_{\alpha\beta}\eta_{\sigma\rho}\eta_{\gamma\lambda})
		  + \frac{1}{2} P_{24}(k_{1}.k_{2}\: \eta_{\mu\alpha}\eta_{\nu\sigma}\eta_{\beta\gamma}\eta_{\rho\lambda}) \right. \\
		  & \left.\phantom{\left[ \right.} + \frac{1}{2}P_{24}(k_{1\nu}k_{1\beta}\eta_{\mu\sigma}\eta_{\alpha\gamma}\eta_{\rho\lambda})
		  +\frac{1}{2}P_{12}(k_{1\sigma}k_{2\gamma}\eta_{\mu\nu}\eta_{\alpha\beta}\eta_{\rho\lambda})
		  + P_{24}(k_{1\nu}k_{2\sigma}\eta_{\beta\mu}\eta_{\alpha\gamma}\eta_{\rho\lambda}) \right. \\
		  & \left.\phantom{\left[ \right.} -P_{12}(k_{1}.k_{2}\: \eta_{\alpha\nu}\eta_{\beta\sigma}\eta_{\gamma\mu}\eta_{\rho\lambda})
		  + P_{12}(k_{1\nu}k_{2\mu}\eta_{\beta\sigma}\eta_{\gamma\alpha}\eta_{\rho\lambda})
		  + P_{12}(k_{1\nu}k_{1\sigma}\eta_{\beta\gamma}\eta_{\mu\alpha}\eta_{\rho\lambda}) \right. \\
		  &\left.\phantom{\left[ \right.} -P_{24}(k_{1}.k_{2} \: \eta_{\mu\alpha}\eta_{\beta\sigma}\eta_{\gamma\rho}\eta_{\lambda\nu})
		  -2P_{12}(k_{1\nu}k_{1\beta}\eta_{\alpha\sigma}\eta_{\gamma\rho}\eta_{\lambda\mu})
		  -2P_{12}(k_{1\sigma}k_{2\gamma}\eta_{\alpha\rho}\eta_{\lambda\nu}\eta_{\beta\mu}) \right. \\
		  &\left.\phantom{\left[ \right.}-2P_{24}(k_{1\nu}k_{2\sigma}\eta_{\beta\rho}\eta_{\lambda\mu}\eta_{\alpha\gamma})
		  -2P_{12}(k_{1\sigma}k_{2\rho}\eta_{\gamma\nu}\eta_{\beta\mu}\eta_{\alpha\lambda})
		  +2P_{6}(k_{1}.k_{2}\: \eta_{\alpha\sigma}\eta_{\gamma\nu}\eta_{\beta\rho}\eta_{\lambda\mu})\right. \\
		  & \left.\phantom{\left[ \right.} -2P_{12}(k_{1\nu}k_{1\sigma}\eta_{\mu\alpha}\eta_{\beta\rho}\eta_{\lambda\gamma})
		  -P_{12}(k_{1}.k_{2}\: \eta_{\mu\sigma}\eta_{\alpha\gamma}\eta_{\nu\rho}\eta_{\beta\lambda})
		  -2P_{12}(k_{1\nu}k_{1\sigma}\eta_{\beta\gamma}\eta_{\mu\rho}\eta_{\alpha\lambda}) \right. \\
		  & \left. \phantom{\left[ \right.} -P_{12}(k_{1\sigma}k_{2\rho}\eta_{\gamma\lambda}\eta_{\mu\nu}\eta_{\alpha\beta})
		  -2P_{24}(k_{1\nu}k_{2\sigma}\eta_{\beta\mu}\eta_{\alpha\rho}\eta_{\lambda\gamma})
		  -2P_{12}(k_{1\nu}k_{2\mu}\eta_{\beta\sigma}\eta_{\gamma\rho}\eta_{\lambda\alpha}) \right. \\
		  & \left. \phantom{\left[ \right.} + 4P_{6}(k_{1}.k_{2} \: \eta_{\alpha\nu}\eta_{\beta\sigma}\eta_{\gamma\rho}\eta_{\lambda\mu})\right]    
     \end{aligned}$
     };
  \end{tikzpicture}
 \end{center}
\end{figure}
where \enquote{sym} stands for symmetrization between $(\mu,\alpha) \: ; \: (\nu,\beta) \: ; (\sigma,\gamma) \: ; (\rho,\lambda)$ and the symbol $ P_{m} $ denotes $m$ number of distinct permutations between the indices $ (k_{1},\mu,\alpha) \: ; \: (k_{2},\nu,\beta) \: ; \: (k_{3},\sigma,\gamma) \: ; \: (k_{4}, \rho,\lambda)$.

\bibliography{ref}
\bibliographystyle{utphys} 

\end{document}